\documentclass[aps,prl,twocolumn,superscriptaddress,showpacs]{revtex4}
\usepackage{graphicx}  
\usepackage{dcolumn}   
\usepackage{bm}        
\usepackage{amssymb}   
\usepackage{amsmath}
\usepackage{verbatim}
\usepackage{color}

\begin{document}

\title{Theory of Absorption Induced Transparency}

\author{Sergio G. Rodrigo} \email{sergut@unizar.es}
\affiliation{Centro Universitario de la Defensa, Ctra. de Huesca s/n, E-50090, Zaragoza, Spain} 
\affiliation{Instituto de Ciencia de Materiales de Arag\'on and
Departamento de F\'isica de la Materia Condensada,CSIC-Universidad
de Zaragoza, E-50009, Zaragoza, Spain}
\author{F.J. Garc\'{i}a-Vidal}
\affiliation{Departamento de F\'isica Te\'orica de la Materia Condensada and 
Condensed Matter Physics Center (IFIMAC), Universidad Aut\'{o}noma de Madrid, Madrid 28049, Spain}
\author{L. Mart\'{i}n-Moreno}
\affiliation{Instituto de Ciencia de Materiales de Arag\'on and
Departamento de F\'isica de la Materia Condensada,CSIC-Universidad
de Zaragoza, E-50009, Zaragoza, Spain}
\date{\today}

\begin{abstract}
Recent experiments (Angew. Chem. Int. Ed. 50, 2085 (2011)) have demonstrated that the optical transmission through an array of subwavelength holes in a metal film can be enhanced by the intentional presence of dyes in the system.
As the transmission maxima occurs spectrally close to the absorption resonances of the dyes, this  phenomenon was christened Absorption Induced Transparency. Here, a theoretical study on Absorption Induced Transparency is presented. The results show that the appearance of transmission maxima requires that the absorbent fills the holes and that it occurs also for single holes. Furthermore, it is shown that the transmission process is non-resonant, being composed by a sequential passage of the EM field through the hole. Finally, the physical origin of the phenomenon is demonstrated to be non-plasmonic, which implies that Absorption Induced Transparency should also occur at  the infrared or Terahertz frequency regimes.
\end{abstract}
 \keywords{}
\maketitle 

The transmission of light through apertures has been attracting the attention of mankind for centuries. 
For instance, it was used in the invention of the camera obscura, it was at the root of the discussions 
on whether light has a corpuscular or wave nature, and nowadays it is central to the technique of optical lithography. 
A limiting factor for many applications is that the transmission process is severely 
impaired if the lateral hole dimension is smaller than the wavelength of light~\cite{BethePhysRev44}. In the last fifteen years, 
several instances have been found in which light transmission through small holes is enhanced:
(i) in arrays of sub-wavelength holes~\cite{EbbesenNature98} and single holes surrounded by surface corrugations~\cite{LezecScience02}, where light scattered 
by each hole, instead of being reflected, couples to surfaces modes which allows the 
build-up of constructive interferences~\cite{MartinMorenoPRL01,LiuNature08}, (ii) through localized 
resonances in single sub-wavelength 
holes~\cite{KoerkampPRL04,DegironOptCommun04}, occurring close to the cutoff frequency of the hole, which 
can be seen either as a zero-order Fabry-Perot resonance~\cite{GarciaVidalPRL05} or transmission in an 
Epsilon-Near-Zero material~\cite{EnghetaPRL06}, and (iii) for Brewster-angle transmission~\cite{AluPRL11}, appearing at 
oblique illumination when the holey surface is impedance matched to the medium of incidence. 
All these phenomena are now generically known as 
Extraordinary Optical Transmission (EOT)~\cite{GarciaVidalRevModPhys09}.
Recently, another case of EOT has been unveiled: Absorption Induced Transparency (AIT)~\cite{HutchisonACI11}. 
Roughly speaking, AIT refers to an enhancement in the transmittance through hole arrays 
that appears when an absorbing
dye is deposited on them. The most intriguing characteristic 
of AIT is that the transmission peak, in the combined hole array plus dye system, appears at the spectral 
position where the bare dye presents 
resonant absorption. Several mechanisms have been proposed, such as dipole-induced dipole interaction between molecules and localized SPPs, cutoff function modification, changes in Fresnel's coefficients~\cite{HutchisonACI11}, and coherent interactions between the hybridized fields of the oscillating elements present in the system~\cite{GarciaPomarOptExp11}, but the origin of AIT still remains unclear.

\begin{figure}[thb!]
\centering\includegraphics[width=7.0cm]{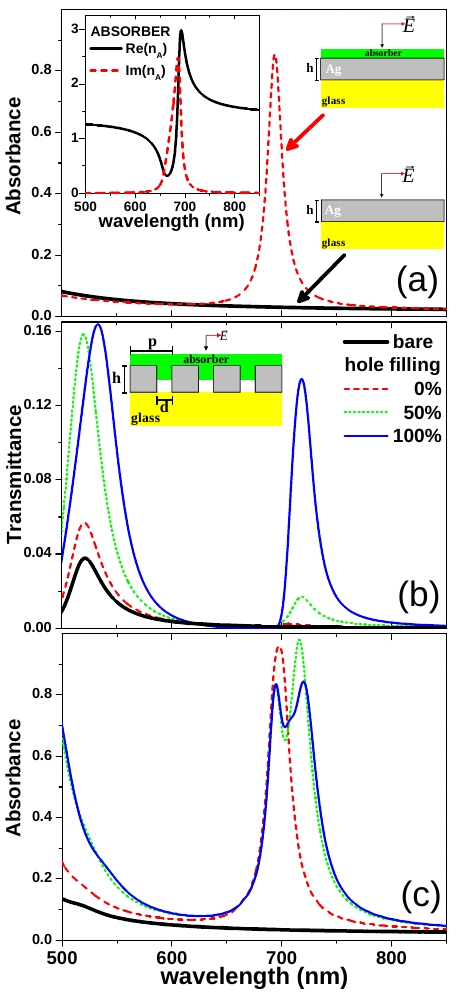} \caption{
(a) Absorption spectra for an optically thick silver film, illuminated at normal incidence from the top interface.
Two configurations are considered: (i) the silver film is coated with an absorbing overlayer (dashed line), 
with the refractive index represented in the inset (see main text for details) and (ii)
no overlayer is present (solid line). (b) Transmittance through a square hole array of circular holes 
on a glass substrate for (i) no absorber present (thick-solid line) and (ii) in the presence of 
an absorbing overlayer and different degrees of hole filling: no filling (dashed line), 
hole half-filled (dotted line) and hole fully-filled (thin-solid line). (c) Absorbance for the same 
configurations as in (b). In all cases, the thickness of the absorbing layer is $30$~nm, 
the thickness of the metal film is $h=200$~nm, the period is  
$p=250$~nm, and the hole diameter is $d=140$~nm.
} \label{fig1}
\end{figure}

In this work we elucidate the physical origin of AIT, which we ascribe mainly to the 
change in the propagation constant inside the holes when filled by the dye. This is originated by the highly dispersive optical response of the dye at frequencies close to the absorption band. 
In particular, AIT transmission maxima very closely coincide with minima in 
the imaginary part of the propagation constant of the hole, and vice versa.
We predict that also single holes should exhibit AIT. We also show that the effect does not have a plasmonic origin and it is thus not restricted to metal films in the optical regime. 

Through this paper the electromagnetic fields (and from them the reflectance, transmittance and absorbance spectra) 
are computed using the Finite Difference Time Domain method~\cite{Taflove05}, and more particularly, 
the implementation to study silver films described in~\cite{RodrigoPRB08}.

Let us first analyze the experimental results on AIT, reported for square arrays of 
circular holes carved on a silver film, placed on a glass substrate, and illuminated at 
normal incidence~\cite{HutchisonACI11}. 
The film thickness is $h=200$~nm, 
and the period $p=250$~nm (see schematics in Fig.~\ref{fig1}(b)). 
These parameters will be kept fixed throughout this work. 
The optical response of the dye (which hereafter will be denoted as the absorber) used in the 
experiments is described by a frequency-dependent dielectric 
constant $\varepsilon_{A}(\omega)$. This classical electrodynamics approach 
is fully justified by the high molecular 
concentration levels and low laser powers used in experiments~\cite{FischerAPL02,ValmorraAPL11}.

In order to implement a 
realistic model with a minimum of material parameters, we consider that $\varepsilon_{A}$ is 
characterized by a single Lorentzian term:
$\varepsilon_{A}(\omega)=\epsilon_h - \Delta \epsilon \, \Omega^2/(\omega^2-\Omega^2+\imath \omega \Gamma)$.
The different parameters are fitted to reproduce the main absorption peak appearing 
when an unperforated film is coated with a $30$~nm 
thin absorber (see Fig. 1 in Ref.~\cite{HutchisonACI11}). We obtain 
$\epsilon_h=1.8$ (host medium), $\Delta \epsilon=0.18$, $\Gamma=0.028$~eV (line width), and 
$\Omega=1.8$~eV (transition frequency). 

Figure~\ref{fig1}(a) shows the absorption spectrum for the considered silver film, with and without the 30~nm 
thick absorbing layer (dashed curve and solid curve, respectively).  As expected, light is mainly reflected 
by the bare flat silver film, with an absorbance spectrum showing a smoothly decreasing trend as the 
wavelength increases. The presence of the thin absorber layer induces a large absorption peak, which 
is correlated with the resonant behavior in the imaginary part of the absorber refractive index, $\mathrm{Im}(n_{A})$, represented in the inset of Fig~\ref{fig1}(a).

Figure~\ref{fig1}(b) renders the transmittance spectra for a hole array (with $140$~nm diameter circular holes)
on a glass substrate, for different configurations of the absorber. The thick solid line 
renders the transmittance for the case when no absorber is present, showing an 
EOT peak at $\lambda \sim 525$~nm. At this wavelength the waveguide modes inside 
the hole are evanescent (the cutoff wavelength for the fundamental mode 
is $\lambda_c^o\sim 430$~nm), and the EOT peak originates from the grating-assisted resonant excitation of 
surface electromagnetic modes at the glass-metal 
interface~\cite{EbbesenNature98,MartinMorenoPRL01} (surface plasmon 
polaritons, SPPs, in the optical regime). 
The other configurations analyzed in Fig.~\ref{fig1}(b) have a $30$~nm absorber thin 
overlayer and different degrees of hole filling: no absorber inside 
the hole (dashed line), absorber filling 50\% of the hole length (dotted line) 
and holes completely filled  (thin solid line). The addition of the absorber induces two 
apparent changes to the transmittance spectra. First, the EOT peak 
at $\lambda \sim 525$~nm is slightly increased by 
the presence of the over layer (due to the stronger binding of the SPP to the surface and 
the corresponding enhancement of the SPP-hole coupling), and more strongly boosted by 
the filling of the holes. In this case, the mechanism behind is the hybridization between SPPs 
and localized resonances~\cite{KoerkampPRL04,DegironJOptA05}, which occurs near the 
cut-off wavelength ~\cite{GarciaVidalPRL05,CarreteroPalaciosPRB12}. As $n_A\approx 1.2$ in 
the wavelength range from $500$~nm to $600$~nm, the localized resonance is expected near 
$\sqrt{n_A}\lambda_c^o\approx 515$~nm, thus overlapping with the SPP. 
The second change is, in agreement with the experimental finding~\cite{HutchisonACI11}, the 
appearance of the AIT peak at $\lambda \sim 720$~nm.  Our calculations show that the intensity of 
the AIT peak strongly depends on the degree of filling of the holes: 
it is negligible when the absorber does not enter into
the hole and maximum when the hole is completely filled. This transmittance peak is spectrally 
located {\it close} to the absorption line of the dye, but it is redshifted by $\sim 25$~nm (in fact,
 a closer inspection reveals that the transmittance has a minimum very close to the 
 maximum of $\mathrm{Im}(n_{A})$). This shift is even more apparent in the absorbance, 
 which is shown in Fig.~\ref{fig1}(c) for the same systems considered in Fig.~\ref{fig1}(b). 
 This magnitude, presents a double-peak structure in the AIT region. One absorption maximum 
 occurs at the same spectral position as that of the absorbing overlayer and, 
 as the transmittance in the hole array is small there, it does not depend much on whether the holes are filled. 
 The other absorption maximum 
 coincides with the transmission peak thus, it only occurs when the holes are at least partially filled. 
 
\begin{figure}[thb!]
\centering\includegraphics[width=\columnwidth]{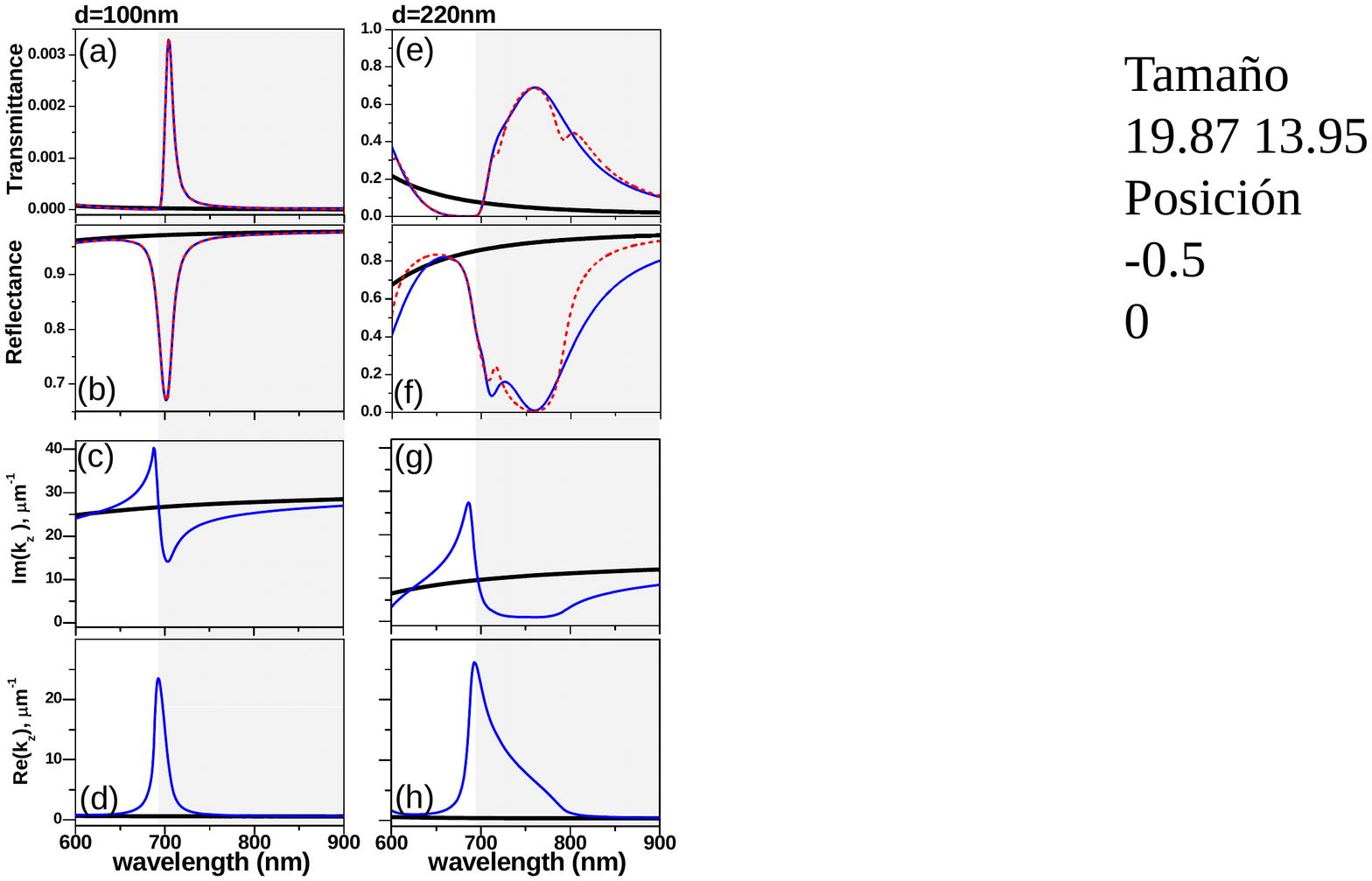}
\caption{Panels (a), (b), (c) and (d) show the transmittance, reflectance, imaginary and 
real part of the propagation constant inside the holes, respectively, for a free standing 
hole array with $p=250$~nm, $h=200$~nm, and  $d=100$~nm. Panels (e), (f), (g) and (h) 
show the corresponding quantities for  $d=220$~nm. Solid thick lines correspond to bare 
structures (without absorber) while thin solid lines are for filled holes. The dashed lines in 
the panels for transmittance and reflectance render the calculations neglecting multiple 
scattering (see main text for details).}
\label{fig2}
\end{figure}

AIT-peaks can be related to the spectral features of the propagation constant
 of filled holes, $k_z$, which has a implicit analytical expression for  circular waveguides~\cite{Jackson75}. 
 This is illustrated in Fig.~\ref{fig2}, where we focus on the AIT spectral region. We
 simultaneously represent the transmittance and reflectance for two hole arrays 
(characterized by different hole diameters) together with the imaginary and real part of $k_z$ inside the holes.
In order to show that the presence of the substrate is not essential, the arrays are considered 
free standing (dielectric constant of substrate equal to unity).  Both period $p=250$~nm and film 
thickness $h=200$~nm are as in the previous figures. In each panel, the thick continuous curves are 
for empty holes while thin ones are for holes filled with the absorber (dashed lines will be 
discussed later on). The correlation between propagation constant and 
scattering coefficients shows that the enhancement of the transmission is due to a 
reduction in the imaginary part of $k_z$, and it indicates that AIT has a localized character. 
Notice also that transmittance peaks are
narrower than reflection dips. This occurs because reflectance decreases when 
transmittance increases, but also when absorption increases, which occurs when the 
imaginary part of $k_z$ is enhanced (anomalous dispersion region). 
\begin{figure}[thb!]
\centering\includegraphics[width=\columnwidth]{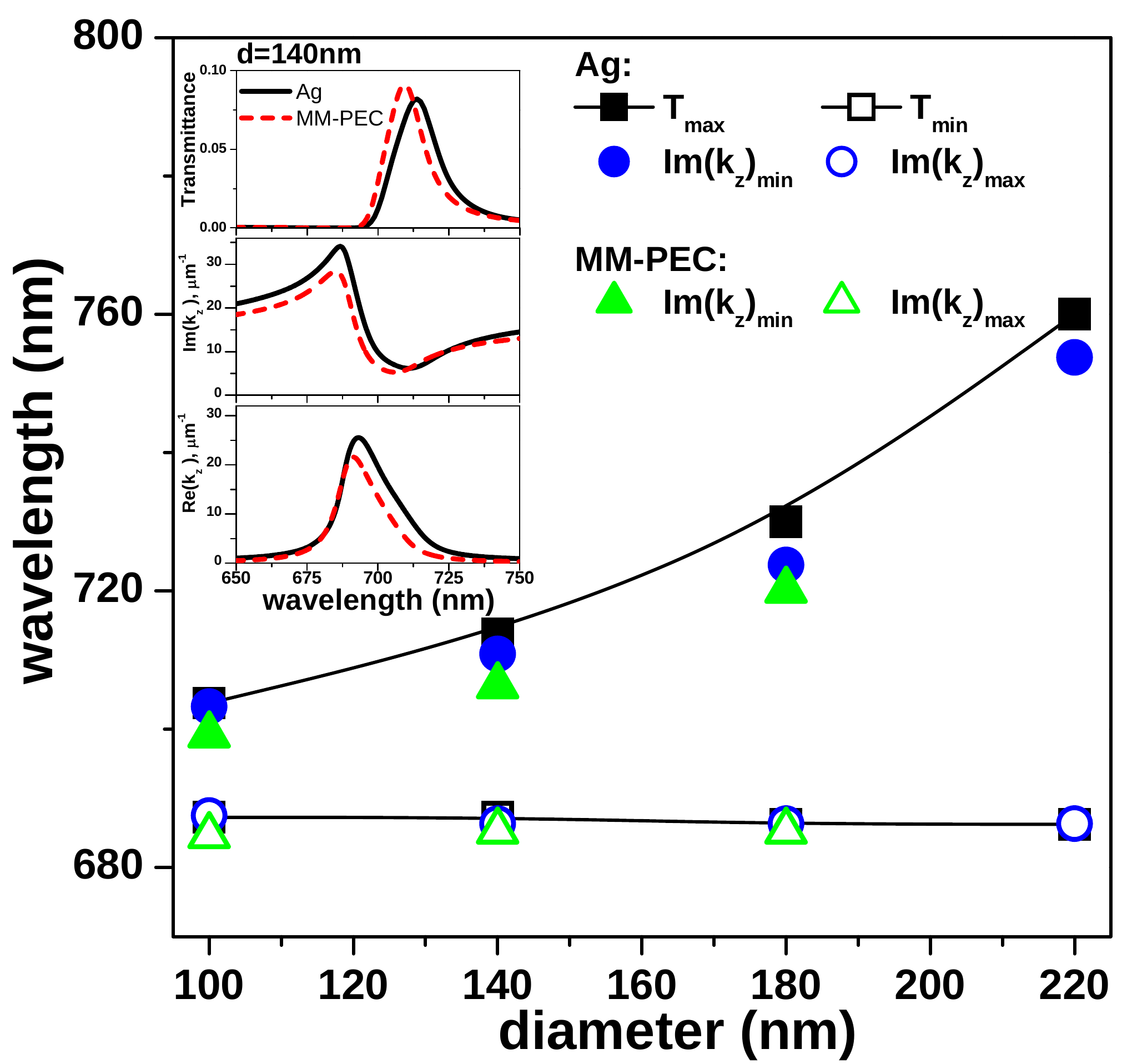}
\caption{Spectral position of AIT transmittance maximum and minimum
 through a free standing hole array in a silver film, as a function of hole diameter (all other geometrical 
 parameters as in Fig.~\ref{fig2}). The transmittance is represented by square symbols, 
 which are filled in the case of maxima and empty for minima. 
Additionally, the spectral position for the maximum of  $Im(k_z)$ is shown with empty circular symbols, 
while that for the minimum is rendered by solid circles. The triangles represent $Im(k_z)$ when the 
metal is considered to be a perfect electrical conductor. Inset: transmittance and $k_z$ for a hole array 
with  $d=140$~nm. The considered metal is silver (solid lines) or PEC (dashed lines). }
\label{fig3}
\end{figure}

A surprising result of the calculations is that AIT is not a resonant effect, at least in the range of 
geometrical parameters considered in the original experiments. This can be demonstrated by 
expressing the zero-order transmission ($t$) and reflection ($r$) amplitude coefficients through the whole 
structure in terms of the partial reflection  coefficient {\it at a single interface} ($\rho$). For a symmetric 
environment, the summation of multiple scattering processes gives  
$t=(1-\rho^2) e^{ik_z h}/(1-\rho^2 e^{2 ik_z h})$ and $r=\rho \, (1-t e^{ik_z h})$. These expressions
have been used in the past to analyze EOT 
phenomena~\cite{MartinMorenoPRL01,deAbajoRevModPhys07,GarciaVidalRevModPhys09}. 
Resonant transmission is typically 
observed when the denominator in $t$ takes values close to zero. Of course, the previous expressions 
also apply to AIT but, in this case the multiple scattering terms can be neglected. This is readily done 
by considering that the reflection process only involves the first interface, so that $r \approx \rho$, while 
the transmission through the array is a three-step process: transmission into the hole, 
propagation inside it and transmission out of the hole, leading to $t \approx (1-\rho^2) e^{ik_z h}$. 
The dashed red lines in the transmittance and reflectance panels in Fig.~\ref{fig2} are obtained 
using this approximation, with the coefficient $\rho$ computed numerically with the FDTD method.
The excellent agreement with the full result confirms that AIT is not a resonant phenomenon in the sense
of requiring the resonant built-up provided by multiple scattering. Instead, AIT transmission can be seen as 
a sequential process, favored by the reduced evanescence inside the holes induced by the absorber. 
This occurs at the long-wavelength side of the absorption resonance, where Kramers-Kronig relations provide 
a combination of large $\mathrm{Re}(n_A)$ and yet relatively small $\mathrm{Im}(n_A)$.

It is possible, following previous work on EOT~\cite{MartinMorenoJoP08,GarciaVidalRevModPhys09} 
and spoof plasmons~\cite{PendryScience04}, to 
develop an analytical minimal model for AIT. 
For that, the electromagnetic fields in different regions of space are expanded in the corresponding 
eigenfunctions, with expansion coefficients that
are fixed by imposing the appropriate boundary conditions. Two approximations are then enforced: (i) only the 
least decaying mode inside the hole and the zero-order diffraction order in the radiation 
regions are considered, 
and (ii) the properties of the metal only enter in that the hole radius is phenomenologically enlarged by 
the actual skin depth~\cite{MartinMorenoOptExp04}; otherwise the metal is represented as a perfect 
electrical conductor (PEC, $\epsilon_{metal}=-\infty$). Within this simple model (which will be 
denoted MM-PEC) 
it is straightforward to show that the 
perforated metal film with filled holes behaves as a {\it uniform} film characterized by a dielectric constant 
 $\tilde{\varepsilon}=(\varepsilon_A-(\lambda/\lambda_c)^2)/S^2$,  and an effective magnetic susceptibility
 $\tilde{\mu}=S^2$. In this expressions $S=(\tilde{d}/p) \sqrt{\pi} / \sqrt{2(\gamma^2-1)}$,
 $\lambda_c= \pi \tilde{d}/\gamma$ 
 is the cut-off wavelength of an unfilled circular waveguide in PEC, 
$\tilde{d}=d + 2\delta$, $\delta \approx \lambda/2 \pi \sqrt{\vert Re(\varepsilon_{Ag}) \vert} \approx 25$~nm 
for silver in the optical regime~\cite{RodrigoPRB08}, and $\gamma = 1.848$.
 The inset to Fig.~\ref{fig3} shows, for an array of holes with diameter $d=140$~nm, the comparison 
 between the exact result and that obtained with the minimal model for transmittance, 
 and also the real and imaginary part of the propagation constant inside the film. The good agreement found 
 for all these quantities validate the minimal model. This agreement is not restricted to the particular 
 diameter considered.
 In Fig.~\ref{fig3} we represent the spectral position for both minimum and maximum AIT 
 transmittance, together with the
 corresponding values for $Im(k_z)$, computed both exactly and within the MM-PEC model. 
 These results confirm the
 correlation between transmittance and $Im(k_z)$ discussed above. It also stresses that, 
 as AIT does not have a plasmonic
origin, it should also appear at shorter frequency regimes (THz, mm...) where the PEC approximation 
for the metal is even more accurate (of course, in this case, the holes should be filled by an absorber 
with an absorption line in the required frequency range).

\begin{figure} 
\centering\includegraphics[width=\columnwidth]{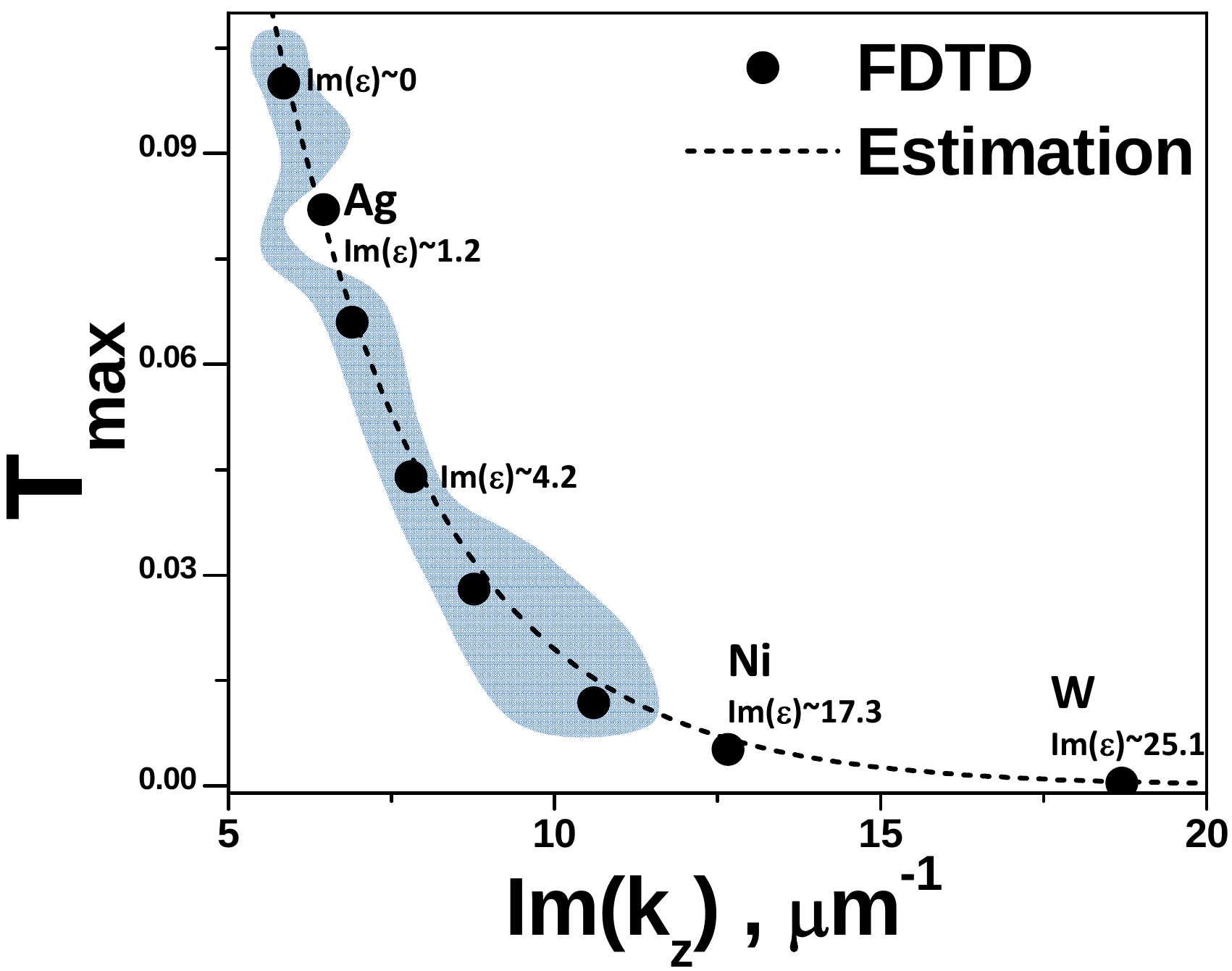}
\caption{Maximum transmittance of the AIT peaks through a hole array of circular holes in a film made 
with different materials (period $p=250$~nm, film thickness $h=200$~nm and hole diameter 
$d=140$~nm) as a function of $Im(k_z)$, for the least evanescent mode inside the waveguide 
 (evaluated at the spectral position of the maximum transmittance). The data points are obtained 
 from FDTD calculations, for films made of either W, Ni or a series of hypothetical metals (points in the shaded area), 
 where the real part of the dielectric constant is essentially that of Ag but the imaginary part is 
 modified. The discontinuous curve is obtained from $T_{max}= C \, \mathrm{exp}(- 2  \mathrm{Im}(k_z) h)$, 
 where the constant $C$ is fixed so that the curve crosses the data point 
 for Ag ($\mathrm{Im}(\epsilon)\sim 1.2$).} \label{fig4}
\end{figure}

It could be argued that the experimentally demonstrated existence of AIT for hole arrays when 
the film is made of Ag 
but not when it is made of either Ni or W~\cite{HutchisonACI11}, 
points to a plasmonic origen.  This is not the case, as AIT is governed by the propagation constant
 inside the hole. This is illustrated in Fig.~\ref{fig4}, that shows the maximum transmittance of the AIT peaks
 through fully filled hole arrays in different metals: Ni, W, and a series of hypothetical metals, 
 where the real part of the dielectric constant is essentially that of Ag (so that the skin depth 
 is practically the same in all the series), but the imaginary part is modified. The transmittance 
 is represented as a function of $Im(k_z)$ for the least evanescent mode inside the waveguide 
 (evaluated at the spectral position of maximum transmittance). The good agreement with 
 the estimation $T_{max}\propto \mathrm{exp}(- 2  \mathrm{Im}(k_z) h)$ (which depends 
 only on the EM properties {\it inside} the hole) shows that AIT in ``bad'' metals is 
 hampered by the enhanced absorption inside the hole, and not by the existence, or lack of it, of 
 surface electromagnetic modes at the metal surface.
 
 \begin{figure} 
\centering\includegraphics[width=\columnwidth]{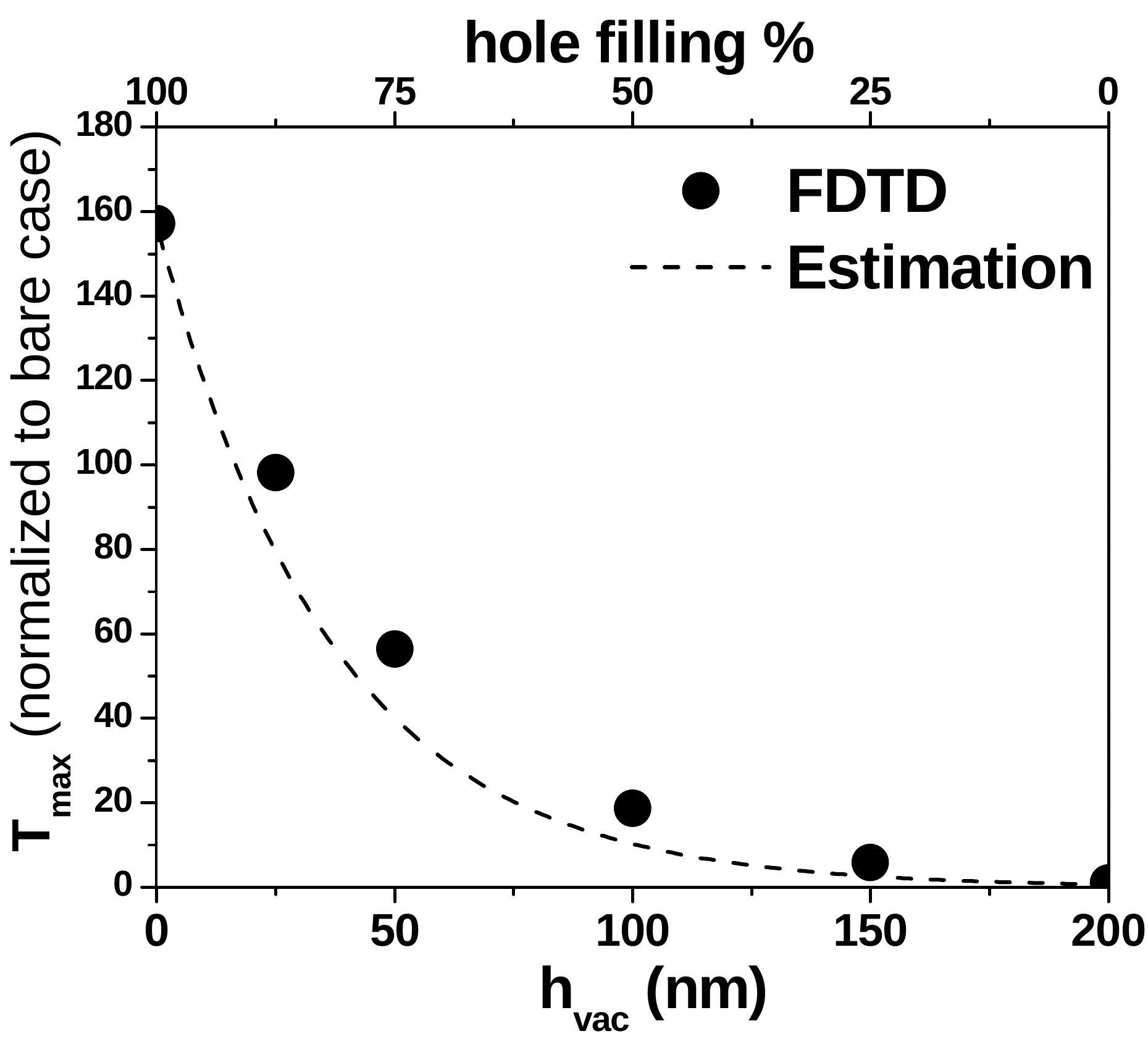}
\caption{Maximum transmittance of the AIT peaks (normalized to the bare case transmitance at the corresponding wavelength) in a hole array of circular holes in silver (period $p=250$~nm, film thickness $h=200$~nm and hole diameter $d=140$~nm).  The holes are filled by the dye a distance $h-h_{vac}$ from the top interface and unfilled a distance $h_{vac}$ from the bottom interface. Solid points represent the results obtained from full FDTD calculations, while the discontinuous line is obtained by a model where the transmission process is assumed to be sequential (see main text for details).} \label{fig5}
\end{figure}

Thus, the main physical mechanism involved in AIT, at least in the range 
of parameters experimentally considered, is the reduction in the opacity of the metal film due to the
filling of the holes with a material with large dielectric constant {\it and} moderate absorption 
(so that multiple scattering inside the hole is not relevant). The asymmetric shapes of the 
transmission peaks are, therefore, due to the spectral asymmetry of the real part of the 
refractive index close to a resonance, and not to any ``Fano-like'' resonant effect.  This
conclusion is further supported by analyzing the maximum transmission for arrays of 
partially-filled holes (where the hole is filled a distance $h-h_{vac}$ from the top surface and unfilled otherwise).
Figure~\ref{fig5} shows, together with the FDTD results, the curve 
$T_{max}(h_{vac}) = T_{max}(0) \,  \mathrm{exp}(- 2  \mathrm{Im}(k_z^{vac}-k_{z}) h_{{vac}})$. 
This expression is obtained by assuming that the transmission through the 
structure is a sequential process, in which the propagation in each region is dominated by 
the corresponding propagation constant ($k_z^{vac}$ in the un-filled hole region and $k_z$ 
in the filled one). Again, the good agreement between the full calculations and this extremely 
simplified model indicates that the main factor governing AIT is the modification of the propagation constant.

\begin{figure} 
\centering\includegraphics[width=\columnwidth]{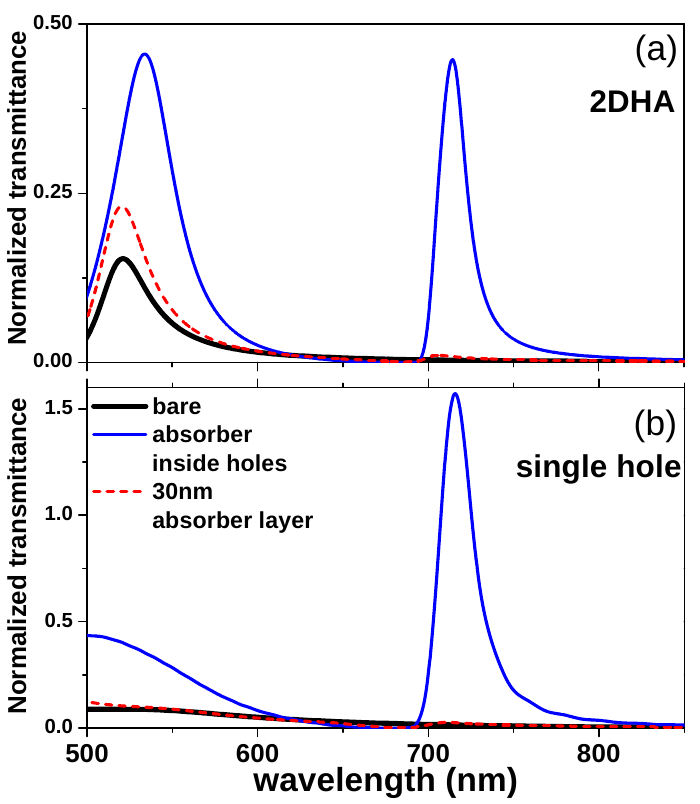}
\caption{(a) Transmittance through a hole array of circular holes in silver (period $p=250$~nm, film 
thickness $h=200$~nm and hole diameter $d=140$~nm).  The thick solid line is for the bare hole array, 
the thin solid line when the holes are filled with an absorber and the thin dashed line when holes are 
empty but the array is coated with a $30$~nm absorbing layer. (b) Transmittance thorough a single 
hole for the same cases as in panel (a). In both cases the transmittance has been normalized to the 
hole area.} \label{fig6}
\end{figure}

It is also interesting to study whether AIT also occurs for isolated holes. 
In Fig.~\ref{fig6}(a) we compare the transmittance through both a hole array (top panel) 
and a single hole (bottom panel). The period, metal thickness, hole diameter and substrate 
are as in the case considered in Fig.~\ref{fig1}. We analyzed three different configurations: 
no absorber, a $30$~nm absorber overlayer with empty holes, and fully filled holes without overlayer.  
These calculations confirm the hybrid nature of the EOT peak appearing at $\sim 525$~nm, 
which for hole arrays already occurs when no dye is present, while in a single hole only occurs 
when the dye fills the hole and induces the presence of localized resonances. Regarding AIT, 
the fact that, for both single holes and hole arrays, the peak emerges only when holes are 
filled confirms that AIT is a localized effect, where collective interactions between holes are not essential.

To conclude, we have demonstrated that AIT requires the presence of absorbers, 
like molecules embedded in a polymer,\textit{ inside}
the holes of a perforated metal film. Our calculations predict that AIT should also occur in single holes,
having thus a localized character.
We have found that their spectral position and width, and their intensity, are 
mainly controlled by the imaginary part of the
propagation constant of holes. 
We have demonstrated that hole arrays in the AIT regime behave like a 
metamaterial, characterized by a dielectric constant composed by a Drude plasma term 
(of geometric origin) plus a Lorentz term arising by the presence of the absorber.
We have shown that AIT peaks are non-plasmonic in character, so they are predicted to occur
in frequencies regimes other than the optical. 
This opens the door for detection spectroscopy of chemical compounds
characterized by sharp absorption lines in the THz or microwave regimes.\\

\begin{acknowledgments}
We acknowledge support from the Spanish Ministry of Science and
Innovation under projects MAT2011-28581-C02, and CSD2007-046-Nanolight.es.
\end{acknowledgments}


\end{document}